\documentclass[prd,preprintnumbers,nofootinbib,superscriptaddress]{revtex4}
\usepackage[dvipdfmx]{graphicx}
\usepackage{enumerate}
\usepackage{amsmath,amssymb}
\usepackage{mathrsfs}
\usepackage{bm}
\usepackage{braket}
\usepackage{ytableau}

\newcommand{\tr}{\operatorname{tr}}
\ytableausetup{boxsize=1em}
\ytableausetup{centertableaux}

\begin{document}

\preprint{CHIBA-EP-223, 2017.06.13}
\title{Double-winding Wilson loops in the SU(N) Yang-Mills theory
}

\author{Ryutaro Matsudo} 
\email{afca3071@chiba-u.jp}
\affiliation{Department of Physics, 
Faculty of Science and Engineering, 
Chiba University, Chiba 263-8522, Japan}

\author{Kei-Ichi Kondo} 
\email{kondok@faculty.chiba-u.jp}
\affiliation{Department of Physics,  
Faculty of Science,
Chiba University, Chiba 263-8522, Japan}




\begin{abstract}
We consider  double-winding, triple-winding and multiple-winding Wilson loops in the $SU(N)$ Yang-Mills gauge theory.
We examine how the area law falloff of the vacuum  expectation value of a multiple-winding Wilson loop depends on the number of color $N$. 
In sharp contrast to the difference-of-areas law recently found for a double-winding $SU(2)$  Wilson loop average, 
we show irrespective of the spacetime dimensionality that a double-winding $SU(3)$ Wilson loop follows a novel area law which is neither difference-of-areas nor sum-of-areas law for the area law falloff 
and that the difference-of-areas law is excluded and the sum-of-areas law is allowed for $SU(N)$ ($N \ge 4$), provided that the string tension obeys the Casimir scaling for the higher representations.
Moreover, we extend these results to arbitrary  multi-winding Wilson loops. 
Finally, we argue that the area law follows a novel law, which is neither sum-of-areas nor difference-of-areas law  when $N\ge 3$.
In fact, such a behavior is exactly derived in the   $SU(N)$ Yang-Mills theory in the two-dimensional spacetime.


\end{abstract}


\maketitle


\section{Introduction}

In this paper  we discuss double-winding, triple-winding and more general multiple-winding Wilson loops \cite{Wilson74} in the $SU(N)$ Yang-Mills gauge theory \cite{YM54}.
The ``double-winding''  Wilson loops consist of the contours which wind once around a loop $C_1$ and once around a loop $C_2$ where the two co-planar loops share one point in common and where $C_2$ lies entirely in the minimal area of $C_1$. 
Recently, the  $SU(2)$ case for the double-winding Wilson loop \cite{GH15} has been investigated to study the mechanism for quark confinement.
See e.g., \cite{Greensite03,KKSS15} for reviews of quark confinement.
It has been found that the area law falloff of the vacuum expectation value (or average) of the double-winding Wilson loop follows a difference-of-areas law \cite{GH15}. 
In this paper we examine  how the area law falloff of  a double-winding, triple-winding and arbitrary multiple-winding Wilson loop averages depend  on the number of color $N$ in the $SU(N)$ Yang-Mills theory.

First, we discuss the case where the two loops $C_1$ and $C_2$ are identical for a double-winding Wilson loop and derive the exact operator relation which relates the double-winding Wilson loop operator in the fundamental representation to a single Wilson loop in the higher dimensional representations depending on $N$.
By taking the average of the relation, we find the relation among the Wilson loop averages.
We find that the difference-of-areas law for the area law falloff of a double-winding Wilson loop average recently claimed for $N=2$ is excluded for $N \ge 3$, provided that the string tension obeys the Casimir scaling \cite{Bali01} for the higher representations.  
We show that a double-winding $SU(3)$ Wilson loop average follows a novel area law which is neither difference-of-areas nor sum-of-areas, while the difference-of-areas is excluded and the sum-of-areas law is allowed for $SU(N)$ ($N \ge 4$), although  the  double-winding $SU(2)$ Wilson loop average is consistent with the difference-of-areas law. 

Next, we extend the analysis to a multiple-winding Wilson loop in the $SU(N)$ Yang-Mills gauge theory. 
We give a physical motivation to consider the multi-winding Wilson loop and give the physical interpretation of the obtained results. This enables us to explain how the $SU(2)$ case is so different from the other cases. 

These results are derived from the group theoretical consideration in the case where all loops are identical.  
In this case, a $m$-times-winding Wilson loop operator in the fundamental representation is rewritten as a linear combination of Wilson loop operators in the higher representations which are distinct from the fundamental representation. 
The results do not depend on the dimensionality of spacetime. 
This provides us with the useful information to analyze the area law falloff of the multiple-winding Wilson loop average.

Finally, we discuss the case where the two loops are distinct for a double-winding Wilson loop.
In this case, we argue that the area law follows a novel law $(N-3)S_2/(N-1) + S_1$ with $S_1$ and $S_2$ ($S_2<S_1$)  being the minimal areas spanned respectively by the loops $C_1$ and $C_2$, which is neither sum-of-areas ($S_1+S_2$) nor difference-of-areas ($S_1-S_2$) law  when $N\ge 3$.
Indeed, we show that this behavior is exactly derived in the $SU(N)$ Yang-Mills theory in the two-dimensional spactime.
These results are consistent with the result obtained recently based on the leading order calculations of the strong-coupling expansion within the framework of the lattice gauge theory \cite{KKS17}, which does not depend on the dimensionality of spacetime. 
The results obtained in this paper will give useful information to investigate the true mechanism for quark confinement, which is to be tackled in the subsequent works. 

This paper is organized as follows.
In section  II and III, we give the main results of this paper with their physical interpretation. 
In section  II, we discuss  the area law falloff for a double-winding Wilson loop for the two identical loops. 
In section III, we extend our analysis to    multiple-winding Wilson loops for the $m$ identical  loops. 
In section IV, we treat a double-winding Wilson loop with two distinct loops in the $SU(N)$ Yang-Mills theory in the two-dimensional spacetime. 
Some of the details of the proofs of the main results are given in Appendices.

\section{Double-winding Wilson loop  with identical loops}

For a single closed loop $C$, the Wilson loop operator in the fundamental representation is defined by
\begin{align}
  W(C ) := \frac{1}{N} {\rm tr}[U_F(C) ] , 
\end{align}
where $U_F(C)$ is the parallel transporter along the loop $C $, i.e., the path-ordered product of the group element along the loop $C$:
\begin{align}
  U_F(C ) := P \exp \left\{ ig \int_{C } dx^\mu \mathscr{A}_\mu \right\} \in G  . 
\end{align}
For two closed loops $C_1$ and $C_2$, a double-winding Wilson loop operator in the fundamental representation is defined by
\begin{align}
  W(C_1 \times C_2) := \frac{1}{N} {\rm tr}[U_F(C_1)U_F(C_2)] . 
\end{align}
See Fig.~1. 

\begin{figure}[tbp]
\begin{center}
\includegraphics[height=6.0cm]{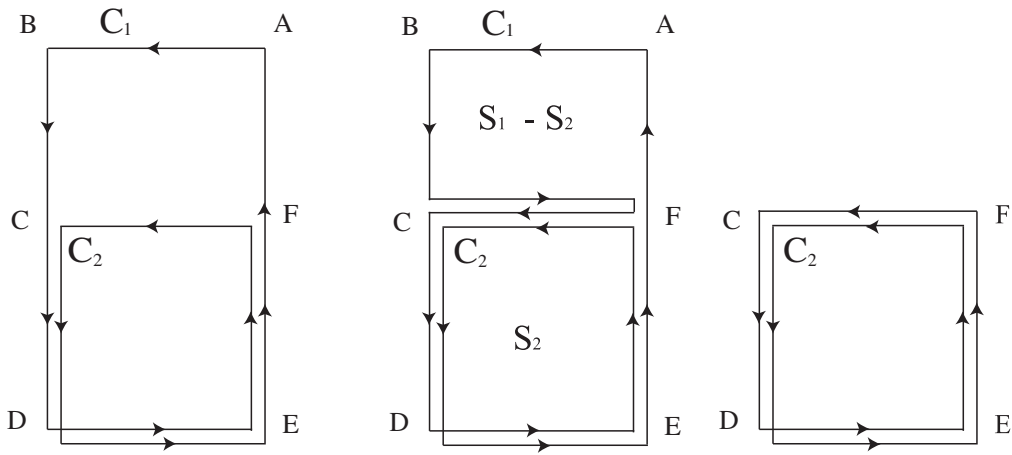}
\end{center} 
\vskip -0.5cm
\caption[]{
The leftmost figure is a double-winding loop with two closed loops $C_1$ and $C_2$ winding in the same direction and the middle one is its deformation. 
The rightmost figure is the case of two identical loops. 
}
\label{C29-fig:W_loop-Sigma}
\end{figure}

In what follows, we consider what type of the area law follows for the double-winding Wilson loop average, irrespective of the lattice and continuum formulations. 
For this purpose, we consider the case of two identical loops, i.e., $C_1=C_2=C$.  
In the identical case, the double-winding Wilson loop operator is written as
\begin{align}
  W(C  \times C ) := \frac{1}{N} {\rm tr}[U_F(C )U_F(C )] . 
\end{align}
The two loops $C_1$ and $C_2$ have the same direction.  The two identical loops correspond to the world line of a pair of quarks in the fundamental representation.
The direct product of two fundamental representations is decomposed into the irreducible representations of the color group $SU(N)$. 

For $SU(2)$, the product of two fundamental representations $\boldsymbol{2}$ is decomposed into a singlet $\boldsymbol{1}$ and a triplet $\boldsymbol{3}$, i.e. adjoint representation:
\begin{align}
 \boldsymbol{2} \otimes \boldsymbol{2} = \boldsymbol{1} \oplus \boldsymbol{3} = \boldsymbol{2} \otimes \boldsymbol{2}^*
\end{align}
Since the color singlet state must not be confined and could be observed,  the string tension must vanish and the area law would disappear. 
In fact, the double-winding Wilson loop operator in the fundamental representation is decomposed into a trivial term and the Wilson loop operator $W(C)_{\rm Adj}$ in the adjoint representation for a single Wilson loop $C$ (see Appendix \ref{sec:double-loop} for the derivation): 
\begin{align}
  W(C \times C) = -\frac12 \boldsymbol{1}  + \frac32 W(C)_{\rm Adj} .  \label{dWil}
\end{align}
This operator identity for the Wilson loops leads to the relation for their averages:
\begin{align}
  \langle W(C \times C) \rangle = -\frac12   + \frac32 \langle  W(C)_{\rm Adj}  \rangle  . 
\end{align}
The adjoint Wilson loop average exhibits the area law in the intermediate distance, since the adjoint quarks are screened by gluons in the long distance.
In the intermediate region, we have  
\begin{align}
  \langle W(C \times C) \rangle = -\frac12 + b_2 e^{-\sigma_{\rm Adj} S } + \cdots  . 
\end{align}
This is consistent with the difference-of-areas behavior and contradicts with the sum-of-areas one, as pointed out by \cite{GH15}. 

The quadratic Casimir operator of a representation with an index $J=\frac12, 1, \frac32, ...$ for $SU(2)$ is given  by
\begin{align}
  C_2(J) =  J(J+1)  , 
\end{align}
which has the specific value for $J=\frac12$ and $J=1$:
\begin{align}
  C_2(\frac12) =  \frac{3}{4} , \quad
  C_2(1) =  2 .  
\end{align}
Suppose that the Casimir scaling for the string tension holds. Then  we find adjoint string tension $\sigma_{\rm Adj}$ is obtained from the fundamental string tension $\sigma_{\rm F}$ using the ratio of the quadratic Casimir operators:
\begin{align}
 \sigma_{\rm Adj}  = \frac{C_2(1)}{C_2(\frac12)} \sigma_{\rm F} = \frac{8}{3}  \sigma_{\rm F}. 
\end{align}

The fundamental representation and its conjugate representation are district in general.
If they happen to coincide, the representation is called a \textit{real representation}.
Otherwise, they are called the \textit{complex representation}.
The group $SU(N)$ allows complex representations for $N \geq 3$.
The $SU(2)$ group is very special, since $\mathbf{2}$ and $\mathbf{2}^*$ are equivalent:
	\begin{align}
		\mathbf{2} = \mathbf{2}^* .
	\end{align}
In $SU(2)$, therefore, 2 quarks $qq$ can make a singlet:
$
		\mathbf{2} \otimes \mathbf{2}^* = \mathbf{2} \otimes \mathbf{2} = \mathbf{3} \oplus \mathbf{1} .
$
Thus, the composite particle $qq=q \bar{q}$ is regraded as a meson $q\bar{q}$ and a baryon $qq$ simultaneously and there is no distinction
between mesons and baryons for $SU(2)$ group.

For $SU(3)$, the product of two fundamental representations $\boldsymbol{3}$ is decomposed into a anti-triplet $\boldsymbol{3}^{\ast}$  and a sextet representation $\boldsymbol{6}$ :
\begin{align}
 \boldsymbol{3} \otimes \boldsymbol{3} = \boldsymbol{3}^{\ast} \oplus   \boldsymbol{6}  .
\end{align}
This is represented as the Young diagram:
\begin{align}
	\ydiagram{1} \otimes \ydiagram{1}  = \ydiagram{1,1} \oplus  \ydiagram{2}  .
\label{Young-diagram1}
\end{align}
For $SU(3)$, there is no color singlet for a pair of two quarks, in sharp contrast with a pair of quark and antiquark where 
\begin{align}
 \boldsymbol{3}\otimes\boldsymbol{3}^{\ast}=\boldsymbol{1} \oplus \boldsymbol{8} = \boldsymbol{3}^{\ast}\otimes\boldsymbol{3} ,
\end{align}
which is represented as the Young diagram:
\begin{align}
	\ydiagram{1} \otimes \ydiagram{1,1}  = \ydiagram{1,1,1} \oplus \ydiagram{2,1}.
\end{align}
In fact, the double-winding Wilson loop operator in the fundamental representation $\boldsymbol{3}$ is decomposed into the Wilson loop $W^*(C)=W(C)_{[0,1]}$ in the (anti)fundamental representation $\boldsymbol{3}^{\ast}$ with the Dynkin indices $[0,1]$ and the Wilson loop operator $W(C)_{[2,0]}$ in the sextet representation $\boldsymbol{6}$ with the Dynkin indices $[2,0]$ (see Appendix  \ref{sec:double-loop}  for the derivation):%
\footnote{
It is also possible to rewrite 
\begin{align}
  2 {\rm tr}(U_{[2,0]}) = {\rm tr}(U_{[1,0]}^2) + ({\rm tr}[U_{[1,0]}])^2     , 
\end{align}
which is equal to 
\begin{align}
    4 W(C)_{[2,0]} = W(C  \times C ) + 3  W(C)_{[1,0]}^2    . 
\end{align}
This operator relation leads to the relation for the average:
\begin{align}
 4 \langle W(C)_{[2,0]} \rangle = \langle W(C  \times C ) \rangle + 3  \langle W(C)_{[1,0]}^2 \rangle    . 
\end{align}
This relation was used to examine the Casimir scaling for the representation $[2,0]$ on the lattice, see eq.(5.15) of \cite{Bali01}.
} 
\begin{align}
  W(C \times C) = - W(C)_{[0,1]} + 2W(C)_{[2,0]} . 
\end{align}
This identity leads to the relation for the average:
\begin{align}
  \langle W(C \times C) \rangle = -\langle   W(C)_{[0,1]}  \rangle  + 2 \langle  W(C)_{[2,0]}  \rangle   . 
\end{align}
In the confinement phase, both Wilson loop averages $\langle   W(C)_{[0,1]}  \rangle$ and $\langle  W(C)_{[2,0]}  \rangle$ exhibit the area law for the loop $C$ of any size larger than a critical size below which the Coulomb like behavior is dominant, since they are not screened by  gluons which belong to the adjoint representation $\boldsymbol{8}$ with the Dynkin indices $[1,1]$.
Therefore, we have  
\begin{align}
  \langle W(C \times C) \rangle = a_3 e^{-\sigma_{\rm F}  S   } + b_3 e^{-\sigma_{[2,0]} S } + \cdots  \quad (a_3 < 0 , b_3 > 0 ). 
\end{align}

In the intermediate region, we assume the Casimir scaling to estimate the string tension $\sigma_{R}$ in the  higher-dimensional representation $R$. 
The dimension of the representation with the Dynkin indices $[m,n]$ for $SU(3)$ is given by
\begin{align}
  D([m,n]) =  \frac12 (m+1) (n+1)(m+ n+2)  . 
\end{align}
The quadratic Casimir operator of the  representation with the Dynkin indices $[m,n]$ for $SU(3)$ is given  by \cite{book-group}
\begin{align}
  C_2([m,n]) =  \frac13 (m^2+mn+n^2) + m+ n  , 
\end{align}
with the specific values:
\begin{align}
  C_2([0,0]) =  0 , \quad
  C_2([1,0]) =  C_2([0,1]) = \frac{4}{3} , \quad
  C_2([2,0]) =  \frac{10}{3} , \quad 
  C_2([1,1]) =  3, ...  . 
\end{align}
Assuming the Casimir scaling for the string tension, therefore, the string tension $\sigma_{[2,0]}$ of the representation $[2,0]$ is obtained as the ratio to the fundamental string tension $\sigma_{\rm F}=\sigma_{[1,0]}$:
\begin{align}
 \sigma_{[2,0]} = \frac{C_2([2,0])}{C_2([1,0])} \sigma_{\rm F} = \frac52  \sigma_{\rm F}. 
\end{align}
Therefore, the area law falloff of the double-winding $SU(3)$ Wilson loop average is given in the intermediate region by
\begin{align}
  \langle W(C \times C) \rangle = a_3 e^{-\sigma_{\rm F}  S   } + b_3 e^{-\frac52  \sigma_{\rm F} S } + \cdots  \quad (a_3 < 0 , b_3 > 0 ). 
\end{align}

In the asymptotic region, on the other hand, the string tension $\sigma_{R}$ for quarks in the representation $R$ is determined only through  the $N$-ality $k$ of the representation $R$ (See e.g., section 10.5 of \cite{KKSS15}). 
Notice that the two representations $\boldsymbol{3}^{\ast}=[0,1]$ and $\boldsymbol{6}=[2,0]$ have the same $N$-ality $k=2$.  
Therefore, the two string tensions $\sigma_{[0,1]}$ and $\sigma_{[2,0]}$  converge to the same asymptotic value which is expected to be the fundamental string tension:
\begin{align}
\sigma_{[0,1]} , \sigma_{[2,0]}   \to \sigma_{\rm F}=\sigma_{[1,0]} .
\end{align}

Thus the area law falloff of the double-winding $SU(3)$ Wilson loop average with two identical loops has the same dominant behavior as that of a single-winding Wilson loop average in the fundamental representation. 
\begin{align}
  \langle W(C \times C) \rangle \simeq  c_3 e^{-\sigma_{\rm F}  S   }  . 
\end{align}
This is not consistent with the difference-of-areas behavior and contradicts also with the sum-of-areas law.

For $SU(N)$ ($N \ge 4$), we have the decomposition: 
	\begin{align}
		\mathbf{N} \otimes \mathbf{N} = \left( \frac{\mathbf{N}^2-\mathbf{N}}{2} \right)_A  \oplus \left( \frac{\mathbf{N}^2+\mathbf{N}}{2} \right)_S ,
		\quad (N \geq 4).
		\label{SUN-decomp}
	\end{align}
The decomposition (\ref{SUN-decomp}) shows that the $N=3$ case is a bit special:
$
		\mathbf{3} \otimes \mathbf{3} = \mathbf{3}^*_A \oplus  \mathbf{6}_S  ,
$
where the antisymmetric part belongs to $\mathbf{3}^*$ (not $\mathbf{3}$).
In any case, the color singlet $\mathbf{1}$ does not occur for $N \ge 3$. 
	This excludes the difference-of-areas law for the double-winding Wilson loop average for $N \ge 3$, because the difference-of-areas law contradicts with this fact in the identical case. 
The difference-of-areas law is possible only when the color singlet $\mathbf{1}$ occurs in the irreducible decomposition of two quarks. 

In fact, the double-winding Wilson loop operator in the fundamental representation $\boldsymbol{N}$ is decomposed into the Wilson loop $W(C)_{[0,1,0,...,0]}$ in the representation $\frac{1}{2}\mathbf{N}(\mathbf{N}-1)$ with the Dynkin indices $[0,1,0,...,0]$ and the Wilson loop operator $W(C)_{[2,0,...,0]}$ in the representation $\frac{1}{2}\mathbf{N}(\mathbf{N}+1)$  with the Dynkin indices $[2,0,...,0]$ (see Appendix  \ref{sec:double-loop}  for the derivation): 
\begin{align}
  W(C \times C) = - \frac{N-1}{2} W (C)_{[0,1,0,...,0]} + \frac{N+1}{2} W(C)_{[2,0,...,0]} . \label{double}
\end{align}
This operator relation leads to the relation for the average:
\begin{align}
  \langle W(C \times C) \rangle = - \frac{N-1}{2} \langle   W (C)_{[0,1,0,...,0]} \rangle  + \frac{N+1}{2}  \langle W(C)_{[2,0,...,0]}   \rangle  . 
\end{align}
The Wilson loop averages $\langle   W(C)_{[0,1,0,...,0]}  \rangle$ and $\langle  W(C)_{[2,0,...,0]}  \rangle$ exhibit the area law string tensions for any size larger than a critical size below which the Coulomb like behavior is dominant, since they are not screened by  gluons which belong to the adjoint representation $\boldsymbol{N^2-1}$ with the Dynkin indices $[1,0,...,0,1]$.
Therefore, we have  
\begin{align}
  \langle W(C \times C) \rangle = a_N e^{-\sigma_{[0,1,...,0]}  S   } + b_N e^{-\sigma_{[2,0,...,0]} S } + \cdots   \quad (a_N<0, b_N>0).
\end{align}

In the intermediate region, we assume the Casimir scaling for the string tension $\sigma_{R}$ in the  higher-dimensional representation $R$. 
It is shown  that the dimension of the representation with the Dynkin indices  $[m_1,...,m_{N-1}]$ for $SU(N)$ is given by \cite{book-group}
\begin{align}
  D([m_1,...,m_{N-1}]) =&  \frac{1}{2! \cdots (N-1)!} (m_1+1) (m_1+ m_2+2) \cdots (m_1+\cdots+m_{N-1}+N-1)
  \nonumber\\&
  \times (m_2+1) (m_2+ m_3+2) \cdots (m_2+\cdots+m_{N-1}+N-2)
  \nonumber\\&
  \times \cdots  \times    (m_{N-2}+m_{N-1}+2)( m_{N-1}+1) . 
\end{align}
and the quadratic Casimir operator of the  representation with the Dynkin indices $[m_1,...,m_{N-1}]$ for $SU(N)$ is given  by \cite{Slansky81}
\begin{align}
  C_2([m_1,...,m_{N-1}]) =  \frac1{2N}\sum_{k=1}^{N-1} [ N(N-k) k m_k + k(N-k) m_k^2 + \sum_{\ell=0}^{k-1} 2\ell (N-k) m_\ell m_k ]  . 
\end{align}
with the specific values:
\begin{align}
  &C_2([0,\cdots,0]) =  0 , \quad
  C_2([1,0,\cdots,0]) = \frac{N^2-1}{2N}, \quad
 C_2([0,1,0,\cdots,0]) = \frac{(N-2)(N+1)}N  ,\notag\\ 
  &C_2([2,0,\cdots,0]) = \frac{(N+2)(N-1)}N  ,  
  ...  . 
\end{align}
Under the Casimir scaling, the area law falloff of the double-winding $SU(N)$ Wilson loop average is described in the intermediate region by 
\begin{align}
  \langle W(C \times C) \rangle =  a_N \exp \left( - 2\frac{ N-2 }{N-1} \sigma_F S \right)  + b_N \exp \left( - 2\frac{ N+2 }{N+1} \sigma_F S \right) +\cdots  \quad (a_N < 0, b_N>0).
\end{align}
Notice that the first term becomes dominant on the right-hand side for large $S$. 

In the asymptotic region, on the other hand, the string tension $\sigma_{R}$ for quarks in the representation $R$ is determined only through  the $N$-ality $k$ of the representation $R$. 
Notice that the two representations $\frac{1}{2}\mathbf{N}(\mathbf{N}-1)=[0,1,0,...,0]$ and $\frac{1}{2}\mathbf{N}(\mathbf{N}+1)=[2,0,...,0]$ have the same $N$-ality $k=2$, since the Young diagram of (\ref{SUN-decomp}) is the same as the $SU(3)$ case (\ref{Young-diagram1}). 
[The $N$-ality of a representation of $SU(N)$ is equal to the number of boxes in the corresponding Young tableaux (mod $N$).]
Therefore, the two string tensions $\sigma_{[0,1,0,...,0]}$ and $\sigma_{[2,0,...,0]}$ converge to the same asymptotic value, i.e.,  $\sigma_k$ with $k=2$:
\begin{align}
\sigma_{[0,1,0,...,0]} , \sigma_{[2,0,...,0]}   \to \sigma_k   \ (k=2).
\end{align}
If we assume the Casimir scaling also for the asymptotic string tension,   
\begin{align}
 \sigma_k = \frac{k(N-k)}{N-1} \sigma_{\rm F} .
\end{align}
then the area-law falloff of the double-winding $SU(N)$ Wilson loop average with two identical loops has the   dominant behavior in the intermediate and asymptotic regions given by
\begin{align}
  \langle W(C \times C) \rangle \simeq   c_N \exp \left( - 2\frac{ N-2 }{N-1} \sigma_F S \right) , 
  \label{d-Wilson-loop}
\end{align}
If we adopt another scaling known as the  \textit{Sine-Law scaling} suggested by MQCD and softly broken $\mathcal{N}=2$ \cite{DS95},
\begin{align}
 \sigma_k = \frac{\sin \frac{\pi k}{N}}{\sin \frac{\pi}{N}} \sigma_{F} ,
\end{align}
then the asymptotic behavior is given by
\begin{align}
  \langle W(C \times C) \rangle \simeq c_N e^{- 2 \cos \frac{\pi}{N} \sigma_F S }  . 
\end{align}
In any case, the result is not consistent with the difference-of-areas behavior and contradicts also with the sum-of-areas law. 
For $N \ge 3$, the area law falloff obeys neither difference-of-areas nor sum-of-areas law. 

In the large $N$ limit, however, the result is consistent with the sum-of-areas law in the intermediate and asymptotic regions:
\begin{align}
  \langle W(C \times C) \rangle \simeq e^{- k\sigma_F S } \ (k=2)  . 
\end{align}
However, this result is interpreted as just coming from the $N$-ality, rather than reflecting the dynamics of the Yang-Mills theory. 

\section{Multiple-winding Wilson loop with identical loops}

We can extend the above considerations for a double-winding Wilson loop to a triple-winding and more general multiple-winding Wilson loops.
 
For $SU(3)$, we introduce a triple-winding Wilson loop. 
In the identical case, the triple-winding Wilson loop average for $SU(3)$ is related to the baryon potential. Baryons are color singlet composite particles to be observed in experiments. Therefore, the baryon potential should be non-confining and the string tension must be zero. 
Indeed, we have
\begin{align}
\mathbf{3} \otimes \mathbf{3} \otimes \mathbf{3} &= (\mathbf{3} \otimes \mathbf{3}) \otimes \mathbf{3} \nonumber\\
&= (\mathbf{3}^*_A \oplus \mathbf{6}_S  ) \otimes \mathbf{3} \nonumber\\
&=  \mathbf{3}^*_A \otimes \mathbf{3} \oplus \mathbf{6}_S \otimes \mathbf{3} 
\nonumber\\
&= \mathbf{1}_A \oplus \mathbf{8}_{\rm MA}  \oplus \mathbf{8}_{\rm MS}  \oplus \mathbf{10}_S .
\label{333}
\end{align}
Thus, we can identify the baryon with  the color singlet $\mathbf{1}_A$:
\begin{equation}
B =\varepsilon_{abc} q^a q^b q^c .
\end{equation}
Thus, for the gauge group $G=SU(3)$, a baryon is constructed from three quarks as the color singlet object. 
Therefore, both baryons and mesons are colorless combinations to be observed, whereas the respective color and the colorful particle as a constituent cannot be observed according to the hypothesis of color confinement.
Thus, the Wilson loop average with a trivial representation is most dominant and does not exhibit the area law, that is to say, string tension is zero. 

For $SU(N)$ ($N \ge 4$), a baryon cannot be constructed from three quarks, since three product does not contain the singlet for $N \ge 4$:
	\begin{align}
		&\mathbf{N} \otimes \mathbf{N} \otimes \mathbf{N} \nonumber\\
		=& \frac{1}{3} \mathbf{N} ( \mathbf{N} + \mathbf{1} )( \mathbf{N} - \mathbf{1} ) \oplus 
		\frac{1}{3} \mathbf{N} ( \mathbf{N} + \mathbf{1} )( \mathbf{N} - \mathbf{1} ) \nonumber\\
		& \oplus \frac{1}{6} \mathbf{N} ( \mathbf{N} + \mathbf{1} )( \mathbf{N} + \mathbf{2} ) \oplus 
		\frac{1}{6} \mathbf{N} ( \mathbf{N} - \mathbf{1} )( \mathbf{N} - \mathbf{2} ) .
	\end{align}
	For $SU(4)$, incidentally, we can check the following results:
	\begin{align}
		\mathbf{4} \otimes \mathbf{4}^* = &\mathbf{15} \oplus \mathbf{1} , \nonumber\\
		\mathbf{4} \otimes \mathbf{4}  = &\mathbf{10}_{S} \oplus \mathbf{6}_{A} , \nonumber\\
		\mathbf{4} \otimes \mathbf{4} \otimes \mathbf{4} &= ( \mathbf{10}_{S} \oplus \mathbf{6}_{A}) \otimes \mathbf{4} \nonumber\\
		&= \mathbf{20}_{\rm MS} \oplus \mathbf{20}_S \oplus \mathbf{20}_{\rm MA} \oplus \mathbf{4}_A .
	\end{align}
For $SU(4)$, a quark-antiquark pair $q\bar{q}$ can form a color singlet, while the three quarks $qqq$ is unable to form a color singlet.
This is because there are $\frac{4\cdot3\cdot2}{3\cdot2\cdot1}=4$ ways of forming a completely antisymmetric wave function using 3 colors
from 4 colors. 
For $SU(N)$, therefore, we need $N$ quarks to make a color singlet:
	\begin{align}
		B=\varepsilon_{a_1 \cdots a_N} q^a_1 \cdots q^a_N, \quad (N\geq 3).
	\end{align}
This is examined by considering $N$-times-winding Wilson loop operator.

In view of these, we consider the general multiple-winding Wilson loop operator of $m$-times-winding loops , $W(C_1 \times C_2 \times ... \times C_m)$.
We show that the $m$-times-winding Wilson loop operator $W(C  \times C  \times ... \times C )=W(C^m)$ in the fundamental representation is written as the linear combination of a single Wilson loop operators $W_{R_\ell }(C)$ in higher representations $R_\ell $ when all loops are identical:
\begin{align}
	 W(C^m) = \sum_{\ell =1}^{\min(m,N)}(-1)^{\ell +1} \frac{D(R_\ell)}N W_{R_\ell }(C),
	 \label{multi-loop1}
\end{align}
where the representation $W_{R_\ell }$ is specified by the Dynkin indices of the representation $R_\ell $:
\begin{align}
	R_\ell  := \begin{cases}
	[m,0,\ldots,0] &\text{for }\ell =1, 
\\
	[\underbrace{m-\ell ,0,\ldots,0,1}_\ell ,0,\ldots,0] &\text{for }\ell =2,\ldots,\min(m,N-1), 
\\
	[m-N,0,\ldots,0] &\text{for } \ell =N,\ m\geq N,
	\end{cases}
	 \label{multi-loop2}
\end{align}
and $D(R_\ell)$ is the dimension of $R_\ell$, i.e.,
\begin{align}
	D(R_\ell) = \frac{(N+m-\ell )!}{m(\ell -1)!(m-\ell )!(N-\ell )!}.
\end{align}
The proof is given in Appendix \ref{sec:nth power case}.
For a given $SU(N)$, especially, the case $m=N$ is an important physical case corresponding to the baryon potential. 

Then we have the relation for the average 
\begin{align}
	 \braket{W(C^m)} = \sum_{\ell =1}^{\min(m,N)}(-1)^{\ell +1} \frac{(N+m-\ell )!}{mN(\ell -1)!(m-\ell )!(N-\ell )!} \braket{W_{R_\ell }(C)} .
\end{align}
Assuming the area law falloff with the string tension obeying the Casimir scaling, therefore, the most dominant term is given by
\begin{align}
	\braket{W(C^m)} \simeq \begin{cases}
		(-1)^{m-1} c_{Nm} \exp\left(-\frac{m(N-m)}{N-1}\sigma_F S\right) &\text{for }m<N,\\
		(-1)^{N-1} c_{Nm} &\text{for }m=N, \\
		(-1)^{N-1} c_{Nm} \exp\left(-\frac{m(m-N)}{N+1}\sigma_F S\right) &\text{for }m>N,
	\end{cases}
\end{align}
where $S$ is the minimal area of the loop and $c_{Nm}$ are positive constants.

In particular, a triple-winding Wilson loop for the $SU(3)$ Yang-Mills theory is written as 
\begin{align}
  \langle W(C^3) \rangle 
= \frac{10}{3} \langle   W (C)_{[3,0]} \rangle  - \frac{8}{3}  \langle W(C)_{[1,1]}   \rangle + \frac{1}{3}  \langle W(C)_{[0,0]}   \rangle 
= \frac{10}{3} \langle   W (C)_{[3,0]} \rangle  - \frac{8}{3}  \langle W(C)_{[1,1]}   \rangle + \frac{1}{3}   
  , 
\end{align}
where we have used $W(C)_{[0,0]}=\boldsymbol{1}$.
This is consistent with (\ref{333}). 
The triple-winding Wilson loop operator is related to the baryonic Wilson loop operator, see e.g., \cite{Creutz85}.

\section{General double-winding Wilson loop}

Finally, we consider a general double-winding Wilson loop where the two loops are distinct. 
In the two-dimensional spacetime we can exactly calculate the double-winding Wilson loop average.
This fact is first demonstrated by Bralic in \cite{Bralic80} for the $U(N)$ gauge theory.  
The exact result for the double-winding Wilson loop average for $U(N)$ is
\begin{align}
	\braket{W(C_1\times C_2)} = \frac{N+1}2 \exp\left[-\frac{\tilde g^2 N}{2} \left( S_1+\frac{N+2}{N}S_2  \right)\right]
	- \frac{N-1}2 \exp\left[-\frac{\tilde g^2 N}{2} \left( S_1+\frac{N-2}{N}S_2 \right)\right],
\end{align}
where $\tilde g$ is the coupling constant in $U(N)$ gauge theory.
Incidentally, the $U(1)$ case reads
\begin{align}
	\braket{W(C_1\times C_2)} =   \exp\left[-\frac{g^2  }{2} \left( S_1+3S_2  \right)\right] ,
\end{align}
which reduces for the identical loops $S_1=S_2$ to
\begin{align}
	\braket{W(C \times C )} =   \exp\left[-\frac{g^2  }{2}(4S) \right] .
\end{align}
Notice that the area law falloff for the double-winding $U(1)$ Wilson loop average in two-dimensional spacetime does not follow the sum-of-areas law.

Fortunately, we can apply this method to the $SU(N)$ gauge theory. Indeed, by replacing the relations among the generators of $U(N)$ by the ones valid for generators $T_A$ of $SU(N)$ ($A=1,...,N^2-1$):
\begin{align}
	\delta^{AB}T_AT_B &= \frac{N^2-1}{2N} \bm1 ,
\notag\\
	\delta^{AB} (T_A)^{\alpha_1}_{\beta_1} (T_B)^{\alpha_2}_{\beta_2} &= \frac12\delta^{\alpha_1}_{\beta_2}\delta^{\alpha_2}_{\beta_1} - \frac1{2N}\delta^{\alpha_1}_{\beta_1}\delta^{\alpha_2}_{\beta_2} ,
\end{align}
we can obtain the exact result for the double-winding Wilson loop average for $SU(N)$:
\begin{align}
	\braket{W(C_1\times C_2)} = \frac{N+1}2 \exp\left[-\frac{g^2}2\frac{N^2-1}{2N}\left( S_1+\frac{N+3}{N+1}S_2  \right)\right]
	- \frac{N-1}2 \exp\left[-\frac{g^2}2\frac{N^2-1}{2N}\left( S_1+\frac{N-3}{N-1}S_2 \right)\right].
\end{align}
In the large $N$ limit, both $U(N)$ and $SU(N)$ cases agree\footnote{
	The agreement occurs if $\tilde g^2 = g^2/2$.
	}
\begin{align}
	\braket{W(C_1\times C_2)} =  (1-\tilde g^2N S_2) \exp\left[-\frac{\tilde g^2N}{2}\left( S_1+ S_2  \right)\right] .
\end{align}
See e.g. \cite{Makeenko02} for the large $N$ result of $SU(N)$ based on the Makeenko-Migdal loop equation. 

In view of these facts, we give a conjecture for the area-law falloff of the double-winding $SU(N)$ Wilson loop average with two loops $C_1, C_2$:
\begin{align}
	\braket{W(C_1\times C_2)} \simeq -c_{N} \exp\left[- \sigma_{F} \left( S_1+\frac{N-3}{N-1}S_2 \right)\right]  \quad (c_N > 0)  .
\end{align}
This follows from the product of the two area law falloffs for an ordinary  single-winding loop with the area $S_1 - S_2$ and a double-winding loop with the identical area $S_2$ obeying (\ref{d-Wilson-loop}):
\begin{align}
	\braket{W(C_1\times C_2)} \simeq  \exp \left[ - \sigma_{F} \left(S_1 - S_2 \right)\right] \times (- c_{N})\exp\left[-  2\frac{ N-2 }{N-1} \sigma_F S_2 \right] \quad (c_N > 0)  .
\end{align}
This is suggested from the middle diagram of Fig.~\ref{C29-fig:W_loop-Sigma}.
This conjecture is consistent with the above considerations for the identical loops $S_1=S_2$ and reduces to the ordinary area law for $S_2=0$. 
We expect that this result holds also in four dimensions. 
Indeed, this leading behavior could hold irrespective of the spacetime dimension, which is also suggested from the strong coupling expansion of the lattice gauge theory \cite{KKS17}.


\begin{acknowledgements}
R. M. was  supported by Grant-in-Aid for JSPS Research Fellow Grant Number 17J04780.
K.-I. K. was  supported by Grant-in-Aid for Scientific Research, JSPS KAKENHI Grant Number (C) No.15K05042.
\end{acknowledgements}

\appendix
\section{Double-winding case: the derivation of Eqs.\ (\ref{dWil}) and (\ref{double})}\label{sec:double-loop}

First we consider the case $N=2$.
Let $U$ be an element of $SU(2)$.
There exists a group element $V$ such that $VUV^{-1}$ is diagonal.
Let this diagonal matrix be $\operatorname{diag}(\exp(i\theta/2),\exp(-i\theta/2))$.
Thus we can write
\begin{align}
	\tr U^2 &= \tr (VUV^{-1})^2 
	 = e^{i\theta} + e^{-i\theta} 
	 = \tr U_{A} - 1
\end{align}
where $U_A$ denotes the adjoint representation of $U$.
Here we have used the adjoint representation of $VUV^{-1}$ is $\operatorname{diag}(\exp(i\theta),1,\exp(-i\theta))$.
Therefore in the case of the gauge group $SU(2)$ the double-winding Wilson loop operator $W(C\times C)$ can be written using the single-winding Wilson loop operator $W_A$ in the adjoint representation as
\begin{align}
	W(C\times C) = \frac32W_A - \frac12 \mathbf{1}.
\end{align}

When the gauge group is $SU(N)$ ($N\geq 3$), we show the double-winding Wilson loop operator $W(C\times C)$ can be written using the higher dimensional representation as
\begin{align}
	W(C\times C) = \frac{N+1}2W_{[2,0,\ldots,0]} - \frac{N-1}2 W_{[0,1,0,\ldots,0]}
\end{align}
by showing
\begin{align}
	\tr U^2 = \tr U_{[2,0,\ldots,0]} - \tr U_{[0,1,0,\ldots,0]}, \label{double2}
\end{align}
where $U$ is an arbitrary element of $SU(N)$.

Before proceeding to the general $N$ case, we consider the $N=3$ case.
As in the $SU(2)$ case, a group element $U$ can be diagonalized.
Let this diagonal matrix be $\exp(i\bm v\cdot \bm H)$,  $\bm v\cdot\bm H := v_1H_1 + v_2H_2$ where $H_1$ and $H_2$ are the Cartan generators and $v_1,v_2\in\mathbb R$.
Therefore the trace of $U^2$ is
\begin{align}
	\tr U^2   = \sum_i \bra{\nu^i}e^{2i\bm v\cdot\bm H}\ket{\nu^i}
	 = e^{2i\bm v\cdot\nu^1} + e^{2i\bm v\cdot\nu^2} + e^{2i\bm v\cdot\nu^3}, \label{su(3)}
\end{align}
where $\nu^1$, $\nu^2$ and $\nu^3$ are the weights of the fundamental representation and $\ket{\nu^i}$ is the normalized state corresponding to $\nu^i$.
To write this as the sum of the traces in higher dimensional representations, we must find the representation which has the weights $2\nu^1$, $2\nu^2$ and $2\nu^3$.
To do this, let us consider the representation corresponding to the Young diagram
\begin{align}
	\ydiagram{2}.
\end{align}
A state in this representation can be obtained by symmetrizing the tensor product of two states in the fundamental representation, that is to say,
\begin{align}
	\ket{\nu^i}\otimes\ket{\nu^j} + \ket{\nu^j}\otimes\ket{\nu^i}  
\end{align}
belongs to this representation.
Therefore the weights of this representation are $2\nu^1$, $2\nu^2$, $2\nu^3$, $\nu^1+\nu^2$, $\nu^1+\nu^3$, $\nu^2+\nu^3$, and the degeneracy of each state is one.
Since the highest weight of this representation is $2\nu^1=2\mu^1$, this representation is $[2,0]$, where $\mu^i$ denotes a fundamental weight%
\footnote{
	The fundamental weights $\mu^i$ are defined as $N-1$ dimensional vectors that satisfy
	\begin{align*}
		\frac{2\mu^i\cdot\alpha^j}{ \alpha^k \cdot \alpha^k } = \delta_{ij},
	\end{align*}
	where $\alpha^j$ are roots of $SU(N)$.
	The highest weight of the representation $[m_1,m_2,\ldots,m_{N-1}]$ is
	\begin{align*}
		\sum_{i=1}^{N-1} m_i \mu^i.
	\end{align*}
	}.
Generally the trace in the representation $R$ can be written as
\begin{align}
	\tr U_R = \sum_\mu d_\mu e^{i\bm v\cdot \mu}, \label{generaltrace}
\end{align}
where the sum is over the weights $\mu$ of the representation $R$ and $d_\mu$ is degeneracy of the weight $\mu$.
Then the trace of $U$ in this representation is
\begin{align}
	\tr U_{[2,0]} = e^{i\bm v\cdot 2\nu^1}
	+e^{i\bm v\cdot 2\nu^2}
	+e^{i\bm v\cdot 2\nu^3}
	+e^{i\bm v\cdot (\nu^1+\nu^2)}
	+e^{i\bm v\cdot (\nu^1+\nu^3)}
	+e^{i\bm v\cdot (\nu^2+\nu^3)}.
\end{align}
Because $\nu^1+\nu^2+\nu^3=0$, the sum of the last three terms is the trace in the complex conjugate of the fundamental representation.
Therefore we obtain
\begin{align}
	\tr U^2 = \tr U_{[2,0]} - \tr U_{[0,1]}.
\end{align}

Now we consider the general $N$ case.
In this case, we can write $VUV^{-1}= \exp(i\bm v\cdot\bm H)$, where $\bm v\cdot\bm H := \bm v_aH_a$, $H_1,\ldots, H_{N-1}$ are the Cartan generators and $v_a \in\mathbb R$.
Therefore
\begin{align}
	\tr U^2   = \sum_i \bra{\nu^i}e^{2i\bm v\cdot\bm H}\ket{\nu^i}
	 = \sum_i e^{2i\bm v\cdot\nu^i}, \label{7}
\end{align}
where $\nu^1,\ldots,\nu^{N-1}$ are the weights of the fundamental representation and $\ket{\nu^i}$ is the normalized state corresponding to $\nu^i$.
From this expression, it turns out that we must find the representation with the doubled weights $2\nu^1,\ldots,2\nu^{N-1}$.
As in the $N=3$ case we consider the representation corresponding to the Young diagram
\begin{align}
	\ydiagram{2}.
\end{align}
A state in this representation can be obtained by symmetrizing the tensor product of two states in the fundamental representation, that is to say,
\begin{align}
	\ket{\nu^i}\otimes\ket{\nu^j} + \ket{\nu^j}\otimes\ket{\nu^i}  
\end{align}
belongs to this representation.
Therefore the weights of this representation are
\begin{align}
	\nu^i + \nu^j \quad (i,j=1,\ldots,N,\quad i<j)
\end{align}
and the degeneracy of each state is one.
Because the highest weight is $2\nu^1 = 2\mu^1$, this representation is $[2,0,\ldots,0]$.
Then the trace in this representation is
\begin{align}
	\tr U_{[2,0,\ldots,0]}  = \sum_{i\leq j}e^{i\bm v\cdot(\nu^i+\nu^j)} 
	 = \tr U^2 + \sum_{i<j}e^{i\bm v\cdot (\nu^i+\nu^j)}. \label{8}
\end{align}
Next let us consider the representation corresponding the Young diagram
\begin{align}
	\ydiagram{1,1}.
\end{align}
A state in this representation can be obtained by antisymmetrizing the tensor product of two states in the fundamental representation, that is to say,
\begin{align}
	\ket{\nu^i}\otimes \ket{\nu^j} - \ket{\nu^j}\otimes \ket{\nu^i}
\end{align}
belongs to this representation.
Therefore the weights of this representation are
\begin{align}
	\nu^i + \nu^j \quad (i,j=1,\ldots,N,\quad i\neq j),
\end{align}
and the degeneracy of each state is one.
Because the highest weight is $\nu^1+\nu^2 = \mu^2$, this representation is $[0,1,0,\ldots,0]$.
Then the trace in this representation is
\begin{align}
	\tr U_{[0,1,0,\ldots,0]} = \sum_{i<j}e^{i\bm v\cdot (\nu^i+\nu^j)} \label{9}
\end{align}
Therefore by subtracting Eq.\ (\ref{9}) from Eq.\ (\ref{8}) we obtain Eq.\ (\ref{double2}).

Since the dimensions of $[2,0,\ldots,0]$ and $[0,1,0,\ldots,0]$ are $N(N+1)/2$ and $N(N-1)/2$ respectively, the double-winding Wilson loop can be written as
\begin{align}
	W(C\times C) = \frac{N+1}2 W_{[2,0,\ldots,0]} - \frac{N-1}2 W_{[0,1,0,\ldots,0]}.
\end{align}

\section{Multiple-winding case: derivation of eq.(\ref{multi-loop1})}
\label{sec:nth power case}

The trace of the $m$th power of $U$ can be written as 
\begin{align}
	\tr U^m = \sum_{\ell=1}^{\min(m,N)} (-1)^{\ell-1} \tr U_{R_\ell} \label{nth1}
\end{align}
where
\begin{align}
	R_l := \begin{cases}
	[m,0,\ldots,0] &\text{for }\ell=1, \\
	[\underbrace{m-\ell,0,\ldots,0,1}_\ell,0,\ldots,0] &\text{for }\ell=2,\ldots,\min(m,N-1), \\
	[m-N,0,\ldots,0] &\text{for } \ell=N,\ m\geq N.
	\end{cases}
\end{align}
By denoting the representations using Young diagram, we can also write it as for $m>N$
\ytableausetup{boxsize=1em}
\begin{align}
	\tr U^m =
	U_{\scalebox{0.5}{
		\ytableaushort{{} {} {\none[\cdot]}{\none[\cdot]}{\none[\cdot]}{}}
	}}
	-
	U_{\scalebox{0.5}{
		\ytableaushort{{} {} {\none[\cdot]}{\none[\cdot]}{\none[\cdot]}{},
		{}}
	}}
	+ \cdots + (-1)^{l-1}
	U_{\scalebox{0.5}{
		\ytableaushort{{} {} {\none[\cdot]}{\none[\cdot]}{\none[\cdot]}{},
		{},
		{\none[\cdot]},
		{\none[\cdot]},
		{\none[\cdot]},
		{}}
	}}
	+ \cdots + (-1)^{m-1}
	U_{\scalebox{0.5}{
	\ytableaushort{{},{},{\none[\cdot]},{\none[\cdot]},{\none[\cdot]},{}}
	}},
\end{align}
where there are $m$ boxes in all diagrams and there are $\ell$ raws in the diagram in $\ell$th term, and for $m\leq N$
\ytableausetup{boxsize=1em}
\begin{align}
	\tr U^m =
	U_{\scalebox{0.5}{
		\ytableaushort{{} {} {\none[\cdot]}{\none[\cdot]}{\none[\cdot]}{}}
	}}
	-
	U_{\scalebox{0.5}{
		\ytableaushort{{} {} {\none[\cdot]}{\none[\cdot]}{\none[\cdot]}{},
		{}}
	}}
	+ \cdots + (-1)^{l-1}
	U_{\scalebox{0.5}{
		\ytableaushort{{} {} {\none[\cdot]}{\none[\cdot]}{\none[\cdot]}{},
		{},
		{\none[\cdot]},
		{\none[\cdot]},
		{\none[\cdot]},
		{}}
	}}
	+ \cdots + (-1)^{N-1}
	U_{\scalebox{0.5}{
		\ytableaushort{{} {} {\none[\cdot]}{\none[\cdot]}{\none[\cdot]}{},
		{},
		{\none[\cdot]},
		{\none[\cdot]},
		{\none[\cdot]},
		{}}
	}},
\end{align}
where there are only $N$ terms.

This is proven as follows.
The trace of the $m$th power of an element of $SU(N)$ can be written as
\begin{align}
	\tr U^m = \sum_ie^{im\bm v\cdot\nu^i}. \label{tracenwinding}
\end{align}
Here $m\nu^1,\ldots,m\nu^{N-1}$ belong to the set of weights of the representation $[m,0,\ldots,0]$ because the highest weight is $m\mu^1=m\nu^1$ and $m\nu^1,\ldots,m\nu^{N-1}$ are related by Weyl reflections.
Therefore as in the second power case the trace of $m$th power of $U$ can be obtained by subtracting the part which contains the weights other than $m\nu^1,\ldots,m\nu^{N-1}$ from the trace of $U_{[m,0,\ldots,0]}$.
The next step is finding the representation which contains the states corresponding to the weights of $[m,0,\ldots,0]$ other than $m\nu^1,\ldots,m\nu^{N-1}$.

To do this, we consider tensor representations.
Let $\ket{i}$ be a vector in the fundamental representation space whose weight is $\nu^i$.
A vector belonging to $m$th tensor power of the fundamental representation space can be written as
\begin{align}
	\ket{i_1}\otimes\ket{i_2}\otimes\cdots\ket{i_m},
\end{align}
and we denote this by
\begin{align}
	\ket{i_1i_2\ldots i_m}.
\end{align}
It is known that an irreducible representation subspace of the tensor product space corresponds to a Young diagram.
We can obtain a state belonging to an irreducible representation subspace as follows.
First put factors of a tensor product in each boxes of Young diagram.
Second symmetrize in the factors in the same raws of the Young diagram.
Lastly antisymmetrize in the factors in the same columns.
The obtaining state belongs to an irreducible representation subspace.
For example let us consider the Young diagram 
\begin{align}
	\ydiagram{2,1} \label{exyng}
\end{align}
and a state $\ket{j_1j_2j_3}$.
Fist put $j_1$, $j_2$ and $j_3$ into the boxes of the diagram as
\begin{align}
	\ytableaushort{{j_1}{j_2},{j_3}}.
\end{align}
By symmetrizing in $j_1$ and $j_2$, we obtain
\begin{align}
	\ket{j_1j_2j_3} + \ket{j_2j_1j_3}.
\end{align}
By antisymmetrizing in $j_1$ and $j_3$, we obtain
\begin{align}
	\ket{j_1j_2j_3} + \ket{j_2j_1j_3} - \ket{j_3j_2j_1} - \ket{j_2j_3j_1}.
\end{align}
This belongs to an irreducible representation subspace.
It is also known that a basis of an irreducible representation subspace corresponds to a set of \emph{semistandard Young tableaux} (see, e.g., \cite{Fulton96}).
A semistandard Young tableau is obtained by filling in the boxes of a Young diagram with numbers which weakly increase along each row and strictly increase down each column.
In fact, if $N=3$, the basis of the representation in the example, 
\begin{align}
	\{ 
	&2\ket{112}-\ket{211}-\ket{121}, \quad
	2\ket{113}-\ket{311}-\ket{131}, \quad
	\ket{122} + \ket{212} - 2\ket{221}, \quad
	2\ket{223}-\ket{322}-\ket{232}, \notag\\
	&\ket{133} + \ket{313} - 2\ket{331}, \quad 
	\ket{233} + \ket{323} - 2\ket{332}, \quad
	\ket{123} + \ket{213} - \ket{321} - \ket{231}, \quad
	\ket{132} + \ket{312} - \ket{231} - \ket{321}
	\}, \label{basis}
\end{align}
corresponds to the set of the semistandard Young tableaux,
\begin{align}
	\{
	\ytableaushort{11,2},\quad
	\ytableaushort{11,3},\quad
	\ytableaushort{12,2},\quad
	\ytableaushort{22,3},\quad
	\ytableaushort{13,3},\quad
	\ytableaushort{23,3},\quad
	\ytableaushort{12,3},\quad
	\ytableaushort{13,2}\}.
\end{align}
The weights of this representation are
\begin{align}
	&2\nu^1+\nu^2,\quad
	2\nu^1+\nu^3,\quad
	2\nu^2+\nu^1,\quad
	2\nu^2+\nu^3,\quad
	2\nu^3+\nu^1,\quad
	2\nu^3+\nu^2,\quad
	\nu^1+\nu^2+\nu^3. \label{weights33}
\end{align}
The weight space with $\nu^1+\nu^2+\nu^3$ is the two-dimensional space whose basis is the set of the last two elements of Eq.\ (\ref{basis}).

Before proceeding the general $m$ and $N$ case, we consider the case $m=3$ and $N=3$.
Let us consider the representation $[3,0]$, which corresponds to the Young diagram
\begin{align}
	\ydiagram{3}.
\end{align}
Since states in this representation are symmetric in the factors of tensor products, the weights are
\begin{align}
	 3\nu^1,\quad  3\nu^2 ,\quad  3\nu^3 ,\quad  2\nu^1+\nu^2 ,\quad  2\nu^1+\nu^3,\quad  2\nu^2+\nu^1,\quad
	 2\nu^2 + \nu^3,\quad 2\nu^3+\nu^1,\quad 2\nu^3+\nu^2,\quad \nu^1+\nu^2+\nu^3,\quad
\end{align}
and the degeneracy of each state is one.
Therefore by using Eq.\ (\ref{generaltrace}) we obtain the trace in this representation as
\ytableausetup{boxsize=0.5em}
\begin{align}
	\tr U_{\ydiagram{3}} &= e^{i\bm v\cdot 3\nu^1} +e^{i\bm v\cdot 3\nu^2} +e^{i\bm v\cdot 3\nu^3} +
	+e^{i\bm v\cdot (2\nu^1+\nu^2)}
	+e^{i\bm v\cdot (2\nu^1+\nu^3)}
	+e^{i\bm v\cdot (2\nu^2+\nu^1)}
	+e^{i\bm v\cdot (2\nu^2+\nu^3)}
	+e^{i\bm v\cdot (2\nu^3+\nu^1)} \notag\\
	&\quad+e^{i\bm v\cdot (2\nu^3+\nu^2)}
	+e^{i\bm v\cdot (\nu^1+\nu^2+\nu^3)}.
\end{align}
Next we consider the representation corresponding to the Young diagram Eq.\ (\ref{exyng}).
By using Eq.\ (\ref{generaltrace}) and the fact that the weights of this representation is Eq.\ (\ref{weights33}), the degeneracy of $\nu^1+\nu^2+\nu^3$ is two, and the degeneracies of other weights are one, we obtain the trace in this representation as
\ytableausetup{boxsize=0.5em}
\begin{align}
	\tr U_{\ydiagram{2,1}} = 
	e^{i\bm v\cdot (2\nu^1+\nu^2)}
	+e^{i\bm v\cdot (2\nu^1+\nu^3)}
	+e^{i\bm v\cdot (2\nu^2+\nu^1)}
	+e^{i\bm v\cdot (2\nu^2+\nu^3)}
	+e^{i\bm v\cdot (2\nu^3+\nu^1)}
	+e^{i\bm v\cdot (2\nu^3+\nu^2)}
	+2e^{i\bm v\cdot (\nu^1+\nu^2+\nu^3)}.
\end{align}
Therefore
\begin{align}
	\tr U_{\ydiagram{3}} - \tr U_{\ydiagram{2,1}} &=
	e^{i\bm v\cdot 3\nu^1}
	+e^{i\bm v\cdot 3\nu^2}
	+e^{i\bm v\cdot 3\nu^3}
	-e^{i\bm v\cdot (\nu^1+\nu^2+\nu^3)} \notag\\
	&=
	\tr U^3
	-1,
\end{align}
where we have used Eq.\ (\ref{tracenwinding}) and $\nu^1+\nu^2+\nu^3 =0$.
By adding the trace of the trivial representation, i.e., one, we obtain
\begin{align}
	\tr U^3 = \tr U_{\ydiagram{3}} - \tr U_{\ydiagram{2,1}} + \tr U_{\ydiagram{1,1,1}}.
\end{align}
where we have used the fact that the Young diagram
\ytableausetup{boxsize=1em}
\begin{align}
	\ydiagram{1,1,1}
\end{align}
corresponds to the trivial representation.
Because of the degeneracy, when $m\geq 3$ we need more than two representations.

\begin{figure}[t]
\centering
\includegraphics[width=0.4\hsize]{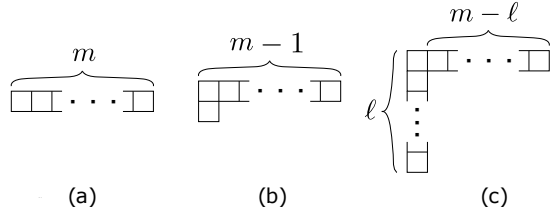}
\caption{Young tableaux associated with $m$-times-winding Wilson loop operator for $SU(N)$ group.}
\label{tableaux}
\end{figure}

Now we consider the general $m$ and $N$ case.
The representation $[m,0,\ldots,0]$ corresponds to the Young diagram shown in Fig.\ \ref{tableaux} (a) because the highest weight of the representation corresponding to the Young diagram is $m\nu^1$, which is the highest weight of $[m,0,\ldots,0]$.
Therefore a weight of $[m,0,\ldots,0]$ can be written as
\begin{align}
	\sum_{\ell=1}^{m} \nu^{i_\ell} \quad (1\leq i_1\leq \cdots\leq i_m\leq N), \label{weights}
\end{align}
and the degeneracy of each weight is one.
This is because the states in this representation can be represented as the symmetric tensor products of $m$ states in the fundamental representation, and there is only one symmetric tensor product which contains $\ket{i_1},\ldots,\ket{i_m}$ as the factors.

Next let us consider the representation $[m-1,1,0,\ldots,0]$, which corresponds to the Young diagram shown in Fig.\ \ref{tableaux} (b).
This representation contains the states which have the weights Eq.\ (\ref{weights}) other than $m\nu^1,\ldots,m\nu^N$ because at least two different states of the fundamental representation must appear as the factors of the tensor products in each state in this representation.
The degeneracy of the weights which have $k$ different weights of the fundamental representation in the sum, i.e.,
\begin{align}
	\sum_{i=1}^k \ell_i\nu^{j_i} \quad ( \ell_i\in\mathbb N,\quad \sum_{i=1}^k\ell_i = m,\quad  1\leq j_1<\cdots<j_k\leq N) 
	\label{weights2}
\end{align}
is $k-1$ 
(notice that $2\leq k\leq \min(m,N)$).
This fact is proven as follows.
The degeneracy of the weights Eq.\ (\ref{weights2}) is the number of the semistandard Young tableaux where the integer $j_i$ appears $\ell_i$ times for $i=1,\ldots,k$.
In the semistandard Young tableaux, $j_1$, which is largest integer in $j_1,\ldots,j_k$, must appear in the first box of the first raw, and since the same number must not appear in the same column, the second box in the first column must be filled by any one of $j_2,\ldots,j_k$.
The entries in the remaining boxes are automatically determined.
This means that the semistandard Young tableau is determined by what is the entry of the second box in the first column.
Thus the number of the corresponding semistandard Young tableau is $k-1$.
Therefore if we subtract $\tr U_{[m-1,1,0,\ldots,0]}$ from $\tr U_{[m,0,\ldots,0]}$, we subtract too much.
We need to consider another representation.

Consider the representation corresponding to the Young diagram shown in Fig.\ \ref{tableaux} (c).
Notice that $\ell\leq m$ and $\ell\leq N$ because there are $m$ boxes in the diagrams and there are no representations corresponding to the Young diagrams which has more than $N$ columns when the group is $SU(N)$.
Since at least $\ell$ different states of the fundamental representation must appear as the factors of the tensor product in each state in this representation, the weights of this representation are Eq.\ (\ref{weights2}) for $k=\ell,\ldots,\min(m,N)$.
The degeneracy of the weights Eq.\ (\ref{weights2}) is ${}_{k-1}C_{\ell-1}$.
This is because, by putting $j_1$ into the first box and $l-1$ of $j_2,\ldots,j_k$ into the boxes in the first column other than first box in ascending order, the numbers which should be put in the remaining boxes are determined and then corresponding semistandard Young tableau is obtained.
This means that the semistandard Young tableau is determined by what is the entry of all boxes except the first one in the first column.
Thus the number of the corresponding semistandard Young tableau is ${}_{k-1}C_{\ell-1}$.
This representation is $R_\ell$ since the highest weight of this representation is $(m-\ell+1)\nu^1+\nu^2+\cdots+\nu^\ell=(m-\ell)\mu^1+\mu^\ell$ for $l<N$, and $(m-N+1)\nu^1+\nu^2+\cdots+\nu^N = (m-N)\nu^1$ for $\ell=N$, where we have used $\nu^1+\cdots+\nu^\ell = \mu^\ell$ for $\ell<N$ and $\nu^1+\cdots+\nu^N=0$.

Because
\begin{align}
	\sum_{\ell=1}^{k} {}_{k-1}C_{\ell-1}(-1)^{\ell-1} = (1-1)^{k-1} = 0,
\end{align}
the contribution from the weights Eq.\ (\ref{weights2}) for $k=2,\ldots,\min(m,N)$ cancels in Eq.\ (\ref{nth1}).
Since Eq.\ (\ref{weights2}) for $k=2,\ldots,\min(m,N)$ is all weights of $[m,0,\ldots,0]$ except $m\nu^1,\ldots,m\nu^N$, Eq.\ (\ref{nth1}) is proven.

By using Eq.\ (\ref{nth1}) we can write the $m$-times-winding Wilson loop operator by using the single-winding Wilson loop operator for the higher dimensional representations:
the $m$-times-winding Wilson loop operator can be written as
\begin{align}
	W(C^m) = \sum_{\ell=1}^{\min(m,N)}(-1)^{\ell-1} \frac{D(R_\ell)}N W_{R_\ell},
\end{align}
where $D(R_\ell)$ is the dimension of $R_\ell$, i.e., 
\begin{align}
	D(R_\ell) = \frac{(N+m-\ell)!}{m(\ell-1)!(m-\ell)!(N-\ell)!}.
\end{align}


\end{document}